%% file: xtej1719.tex
\documentclass[useAMS,usenatbib]{mn2e}
\usepackage{graphicx}                                            
\usepackage{times}
\usepackage{epsfig}
\usepackage{rotating,amssymb}

\include{def_bibtex}
\include{definitions}

\def\simlt{\mathrel{\rlap{\lower 3pt\hbox{$\sim$}}
        \raise 2.0pt\hbox{$<$}}}

 \title[Softening in the new X-ray transient XTE J1719--291]{X-ray softening in the new X-ray transient XTE J1719--291 during its 2008 outburst decay}

   \author[M.~Armas Padilla et al.]{M.~Armas Padilla$^{1}$ \thanks{E-mail:M.Armaspadilla@uva.nl}, N.~Degenaar$^{1}$, A.~Patruno$^{1}$, D.~M.~Russell$^{1}$, M.~Linares$^{2}$, 
 \newauthor T.J.~Maccarone$^{3}$, J.~Homan$^{2}$ and R.~Wijnands$^{1}$\\
$^{1}$Astronomical Institute "Anton Pannekoek", University of Amsterdam, Science Park 904, 1098 XH, Amsterdam, The Netherlands\\
$^{2}$MIT Kavli Institute for Astrophysics and Space Research, 70 Vassar Street, Cambridge, MA 02139, USA\\
$^{3}$School of Physics and Astronomy, University of Southampton, Hampshire SO17 1BJ,United Kingdom}
\begin{document}

\maketitle

\begin{abstract}

The X-ray transient XTE J1719--291 was discovered with
\textit{RXTE/PCA} during its outburst in 2008 March, which lasted at
least 46 days. Its 2-10 keV peak luminosity is 7~$\times$~$10^{35}$
erg~s$^{-1}$ assuming a distance of 8 kpc, which classifies the system
as a very faint X-ray transient. The outburst was monitored with
\textit{Swift}, \textit{RXTE}, \textit{Chandra} and
\textit{XMM-Newton}. We analysed the X-ray spectral evolution during
the outburst. We fitted the overall data with a simple power-law model
corrected for absorption and found that the spectrum softened
with decreasing luminosity. However, the \textit{XMM-Newton} spectrum
can not be fitted with a simple one-component model, but it can be fit
with a thermal component (black body or disc black body) plus power-law
model affected by absorption. Therefore, the softening of the X-ray
spectrum with decreasing X-ray luminosity might be due to a change in
photon index or alternatively it might be due to a change in the
properties of the soft component. Assuming that the system is an X-ray
binary, we estimated a long-term time-averaged mass accretion rate of
\Macclong $ \sim $ 7.7 $\times$ $10^{-13}$ $\sol$ yr$^{-1}$ for a
neutron star as compact object and \Macclong $ \sim $ 3.7 $\times$
$10^{-13}$ $\sol$ yr$^{-1}$ in the case of a black hole. Although no
conclusive evidence is available about the nature of the accretor,
based on the X-ray/optical luminosity ratio we tentatively suggest that
a neutron star is present in this system.

\end{abstract}


\section{Introduction}

X-ray transients spend most of the time in a dim quiescent state, with an X-ray luminosity of $10^{31-34}$ erg s$^{-1}$. It is mostly during outbursts that these systems are discovered, when the luminosity increases by more than two orders of magnitude. The nature of these X-ray transients is varied. Many of them are compact objects (black holes or neutron stars) accreting matter from a companion star. In these systems the outbursts are attributed to a strong increase in the accretion rate onto the compact object due to a hydrogen ionization instability in the accretion disc \citep{Lasota2001}.

The peak luminosity reached during these accretion outbursts ($L^{peak}_{X}$; 2-10 keV) covers a wide range, from $10^{34}$ to $10^{39}$ erg s$^{-1}$. Depending on this luminosity, X-ray transients can be classified as \textit{bright} ($L^{peak}_{X}\sim10^{37-39}$), \textit{faint} ($L^{peak}_{X}\sim10^{36-37}$) or \textit{very faint} ($L^{peak}_{X}\sim10^{34-36}$; see \citealt{wijnands2006}). This classification is not strict since hybrid systems do exist which exhibit large variations in their peak $L_{X}$ from outburst to outburst (e.g. SAX J1747.0-2853; \citealt{Werner2004}; \citealt{Wijnands2002}). \\

Very faint X-ray transients (VFXTs) have been discovered in the last decade thanks to the improvement in sensitivity of X-ray instruments. Currently several tens of VFXTs are known, but despite the reasonable number of sources detected, only very few of them have been studied in detail during outbursts. Hence the characteristics of these peculiar sources as well as their nature are still poorly understood. Some of them are neutron stars accreting from, most likely, low-mass stars, since these systems have shown Type-I bursts (e.g. \citealt{Cornelisse2002}; \citealt{DelSanto2007}; \citealt{Chelovekov2007}; \citealt{Degenaar2009}). Classical novae are a possible class of these very faint transients too. \citet{Mukai2008} have argued that those systems can be a small part of the X-ray transients population in the Galactic center, since they can reach peak luminosities in the 10$^{34-35}$ erg s$^{-1}$ range through nuclear fusion of the matter accreted on the white dwarf surface. Another possibility are the symbiotic X-ray binaries, a small sub-class of low mass X-ray binaries (LMXBs) in which the compact primary, most likely a neutron star, is accreting matter from the wind of an M-type giant companion. Only a few such symbiotic X-ray binaries have been identified so far (e.g., \citealt{Masetti2007}). Also several strongly magnetized neutron stars (B$\sim$10$^{14-15}G$, magnetars) have shown X-ray outbursts with peak luminosities of  $\sim$ 10$^{35}$ ergs s$^{-1}$ (e.g., \citealt{Ibrahim2004}; \citealt{Muno2007}). They are the only known non-accreting systems that can exhibit VFXT outbursts. The cause of these transient outbursts is not fully understood, but it likely is related to a decay in the strong magnetic field of the neutron star \citep{Ibrahim2004}. It is also possible that a fraction of these under-luminous transients are high mass X-ray binaries (HMXBs), i.e., compact objects accreting from a circumstellar disc or the strong stellar wind of a star with a mass higher than 10$\sol$ (e.g., \citealt{Okazaki2001}).\\
 
The low luminosities of VFXTs in which a compact object accretes from low-mass donor in combination with duty cycles of $\lesssim$10\% (as is common for the brighter X-ray transients) imply that the mean accretion rates in these systems are very low (e.g. \citealt{Degenaar2009}). Therefore, such VFXTs provide us with new regimes to study accretion onto compact objects. For example, by studying  the outburst properties of the systems that harbour a neutron star (e.g. displaying X-ray pulsations or bursts) new ways of studying ultra-dense matter can be performed \citep{Wijnands2008}. Moreover, VFXTs yield new inputs for the outburst and evolution models that were developed to explain the bright systems, but are not able to account for all the VFXT manifestations (e.g., \citealt{King2006}). \\

In this work we present an extensive X-ray analysis of XTE J1719-291, which was discovered with Rossi X-ray Timing Explorer/Proportional Counter Array (\textit{RXTE}/PCA) bulge scans on 2008 March 21 \citep{Markwardt2008}. Its outburst was monitored with {\it Swift}, which initially showed an X-ray flux decrease (\citealt{Markwardt2008}, \citealt{Degenaar2008}) but then rebrightened \citep{Degenaar2008b}. A duration of 46 days elapsed between the source's discovery and the time when it was no longer detectable \citep{Degenaar2008a}. The most accurate position was obtained with \textit{Chandra}, $\alpha$= 17h 19m 17.18s, $\delta$= -29d 04' 10.0" with an uncertainty of 0.2" (J2000, 90$\%$ confidence; \citealt{Greiner2008}). At this position a source was detected with the MPI/ESO 2.2m telescope at La Silla, which likely represents the optical counterpart. In a second optical pointing performed 24 days later, the source was not detected \citep{Greiner2008}. From the upper limits during this observation and assuming a distance of 8 kpc an absolute V magnitude of \textgreater 5.8 was derived, which suggests a companion star K0V or later \citep{Greiner2008}.\\

\section{Observations and analysis}

XTE J1719-291 was observed over a sixty day time-span with \textit{RXTE}, {\it Chandra}, {\it XMM-Newton} and {\it Swift} between 2008 March 24 and 2008 May 14. Altogether nine pointed observations were carried out, six observations with {\it Swift}/XRT, one with {\it XMM-Newton}/EPIC, one with {\it Chandra}/HRC and one with \textit{RXTE}/PCA.  A log of the different observations is given in Table \ref{t1}. 
Apart from these pointed observations, we obtained from the literature four additional flux measurements from \textit{RXTE}/PCA scans made in the period of 2008 March 15-25 \citep{Markwardt2008}. 

\begin{table*}
\begin{center}
\caption[]{Log of the observations of XTE J1719-291.}
\label{t1}
\begin{tabular}{cccccc}
\hline 
\hline
Observation & Date & MJD (UTC) & Exposure (ks) & Instrument\\
			& and start time (UT) & & 				&			\\
   
\hline
1 & 2008-03-24 03:38& 54549.152 & 1.7 & \textit{RXTE}/PCA\\
2 & 2008-03-30 08:54& 54555.371 & 44.7 & \textit{XMM}/EPIC\\
3 & 2008-03-30 12:27& 54555.519 & 5 & \textit{Swift}/XRT\\
4 & 2008-04-03 00:07 &54559.005 & 2.5 & \textit{Swift}/XRT\\
5 & 2008-04-09 13:27& 54565.561 & 1.9 & \textit{Swift}/XRT\\
6 & 2008-04-16 15:28& 54572.645 & 1.7 & \textit{Swift}/XRT\\
7 & 2008-04-27 18:23& 54583.766 & 2.2 & \textit{CHANDRA}/HRC-I\\
8 & 2008-04-30 00:44& 54586.031 & 1.9 & \textit{Swift}/XRT\\
9 & 2008-05-14 13:20& 54600.556 & 1.2 & \textit{Swift}/XRT\\

\hline
 
\end{tabular}
\end{center}
\end{table*}

\subsection{\textit{RXTE} data}

We analysed the {\it RXTE} observation of XTE~J1719--291 taken on March 24, 2008. We extracted a spectrum from the proportional counter array (including PCU2 only), using Standard~2 data of all layers. The background was estimated using {\ttfamily pcabackest} (v. 3.6) and the faint source model. A response matrix was created using {\ttfamily pcarsp} (v. 10.1), taking into account the $\sim$0.2~degree offset between the {\it RXTE}  pointing and XTE~J1719--291. We grouped the resulting spectrum to have a minimum of 20 counts per energy bin and applied a systematic error of 1\%.\\

\subsection{\textit{XMM-Newton} data}

XTE J1719-291 was observed with {\it XMM-Newton} on 2008 March 30, with an exposure time of 44 ks. The data were taken using the EPIC detectors, the two MOS and the pn CCD cameras, operated in full window mode with the medium and thick optical blocking filter, respectively. The data were processed with the standard  {\it XMM-Newton} Science Analysis System (SAS v.9.0) to obtain calibrated event lists and scientific products. The observation was affected by a strong background flare. We exclude the data where the count rate  exceeded 1 and 0.5 counts~s$^{-1}$ for the pn and MOS data, respectively, which results in a total live time of 17~ks. The extraction of spectra were carried out using the {\ttfamily xmmselect} task, as well as the associated response matrices (RMF) and the ancillary response files (ARF) using the standard analysis threads\footnote{See http://xmm.esac.esa.int/sas/current/documentation/threads/}. The spectra were grouped to contain 20 counts per bin using the {\tt FTOOL grppha}. Finally, we checked that the data were not affected by pile-up using the SAS task {\ttfamily epatplot}.\\

\subsection{\textit{Swift} data}

Six observations were carried out with the XRT. All data were collected in Photon Counting (PC) mode. The data were processed running the {\ttfamily xrtpipeline} task in which standard event grades of 0--12 were selected. For every observation, spectra, lightcurves and images were obtained with the {\ttfamily Xselect} (v.2.3) package. Source spectra were extracted from a circular region with a radius of 17 pixels. For the background, three circular regions of similar size as the source region were used over nearby source-free regions. The spectra were grouped to have a minimum of 5 counts per energy bin with {\tt grppha}.\\
The spectra were corrected for the fractional exposure loss due to bad columns on the CCD. For this, we created exposure maps with the {\tt xrtexpomap} task, which is used as input to generate ARF with the {\tt xrtmkarf} task. For the RMF the latest version was used from HEASARC calibration database (v.11). \\
Observations 5 and 6 (see Table 1) have the highest count rates (0.5-0.7 counts/sec) and might be affected by pile-up. To test this, we have used the software {\tt ximage}\footnote{See http://www.swift.ac.uk/pileupthread.shtml for \textit{Swift} pile-up thread}. We compared the point spread function of the data with that expected for the XRT, and we found there was no pile-up. \\  
In the last XRT observation (Obs. 9) the source was not detected. The upper limit on the flux was calculated with WebPIMMS HEARSAC tool\footnote{Available from http://heasarc.gsfc.nasa.gov/Tools/w3pimms.html}. An absorbed power-law model with a photon index of 2.74 and a hydrogen column density (\Nh) of 0.53$\times 10^{22}$ cm$^{-2}$ (see section 3) was assumed, and the count rate was calculated using the prescription for small numbers of counts given by \citet{Gehrels1986}.\\

\subsection{\textit{Chandra} data} 

The {\it Chandra} observation was performed using the High Resolution Camera (HRC-I) on 2008 April 27 for an exposure time of 2.1~ks (see also \citealt{Greiner2008}). We obtained these data from the {\it Chandra} data archive. The intrinsic energy resolution of the HRC-I is poor so no spectral fitting can be carried out.\\
Data reduction was performed using the Chandra Interactive Analysis Software (CIAO  v 4.1). We calculated the net source count with {\tt dmextract} task over a circular region with a radius of 12 pixels and the background was taken with an annulus around the source (of 56 pix inner radius, of 98 pix outer radius). The flux was calculated with WebPIMMS HEARSAC tool assuming a power-law model with a photon index of 2.74 and a hydrogen column density (\Nh) of 0.53$\times 10^{22}$ cm$^{-2}$ (see section 3).\\

\begin{table*}
\begin{center}
\caption[]{Spectral results for XTE J1719-291. }
\label{t2}
\begin{tabular}{cc|ccc|cccc}
\hline 
\hline
  & & \multicolumn{3}{c|}{(0.5-10 keV)}&  &\multicolumn{3}{c}{(2-10 keV)}\\
\cline{3-5}
\cline{7-9}
Obs  & $\Gamma$ & \Fx$_{,abs}\ ^{a}$ & \Fx$_{,unabs}\ ^{a}$ & $\lx\ ^{b}$&  &\Fx$_{,abs}\ ^{a}$ & \Fx$_{,unabs}\ ^{a}$ & $\lx\ ^{b}$ \\
   
\hline
&&&&&&&\\
1 & $2.02 \pm 0.08$& $ 112\pm 11$ & $173 \pm 11$ & $133 \pm 8$& & $ 86.8 \pm 10.0$ & $92 \pm 11$ & $70 \pm 8$ \\
2  &$2.74 \pm 0.05 $& $ 2.71 \pm 0.13 $ & $6.21 \pm 0.1$ & $4.75 \pm 0.07$&& $ 1.62^{+0.12}_{-0.11}$ & $1.72 \pm 0.12$ & $1.33 \pm 0.09$ \\
3 & $2.83 \pm 0.25 $& $ 1.93 ^{+ 0.47 }_{-0.36 }$ & $4.7^{+0.3}_{-0.2 }$ & $3.6^{+ 0.3}_{-0.2 }$&& $ 1.1^{+0.4}_{-0.3}$ & $1.19 ^{+0.44 }_{-0.33 }$ & $0.91 ^{+0.32}_{-0.25}$ \\
4 & $2.6 \pm 0.4$& $ 1.89 ^{+0.92 }_{-0.58}$ & $3.97^{+0.81 }_{- 0.41 }$ & $3.04^{+ 0.62 }_{- 0.31 }$&& $ 1.19^{+0.85}_{-0.51}$ & $1.28 ^{+0.89 }_{-0.54}$ & $0.98 ^{+0.68}_{-0.41}$ \\
5 & $2.32 \pm 0.11 $& $ 20.5 ^{+ 2.7}_{-2.3 }$ & $36.5^{+2.5 }_{- 2.1 }$ & $27.9^{+ 1.9 }_{- 1.6 }$&& $ 14.4^{+2.5}_{-2.1}$ & $15.3 ^{+2.6 }_{-2.2 }$ & $11.7 ^{+2.0}_{-1.7}$ \\
6 & $2.15 \pm 0.09 $& $ 34.9 ^{+ 4.0 }_{-3.5}$ & $57.2^{+3.9 }_{- 3.3 }$ & $43.8^{+ 2.9 }_{- 2.5 }$&& $ 25.9^{+3.8}_{-3.3}$ & $27.4 ^{+3.9 }_{-3.4 }$ & $20 \pm 3$ \\
7 & 2.74 (fix) & $ 2.57 \pm 0.15$ & $5.75^{+0.34 }_{- 0.33 }$ & $4.4 \pm 0.3$&& $ 1.5 \pm 0.1 $& $1.61 ^{+0.10 }_{-0.09 }$ & $1.23 \pm 0.07$ \\
8 & $2.7 \pm 0.4$& $ 2.23 ^{+ 1.05 }_{-0.66 }$ & $4.95^{+0.89}_{- 0.40 }$ & $3.79^{+ 0.69}_{-0.31 }$&& $ 1.36^{+0.95}_{-0.57}$ & $1.46 ^{+1.00 }_{-0.60 }$ & $1.12 ^{+0.76}_{-0.46}$ \\
9 & 2.74 (fix)& $<$0.05 & $<$0.11& $<$0.08& &  $<$0.03& $<$0.03 & $<$0.02\\

&&&&&&&\\
\hline
\end{tabular}

\begin{list}{}{}
\item Note.- \Nh has been fixed to 0.53 $\times 10^{22}$ cm$^{-2}$, the value obtained from the \textit{XMM-Newton} power-law fitting (see section 3) .
\item $^{a}$ Flux in units of $10^{-12}$ erg cm$^{-2}$ s$^{-1}$.
\item $^{b}$ X-ray luminosity in units of $10^{34}$ erg s$^{-1}$ calculated from the unabsorbed flux by adopting a distance of 8 kpc.
\end{list}

\end{center}
\end{table*}

\section{Results}

\begin{figure}
\begin{center}
\includegraphics[angle=-90,width=\columnwidth]{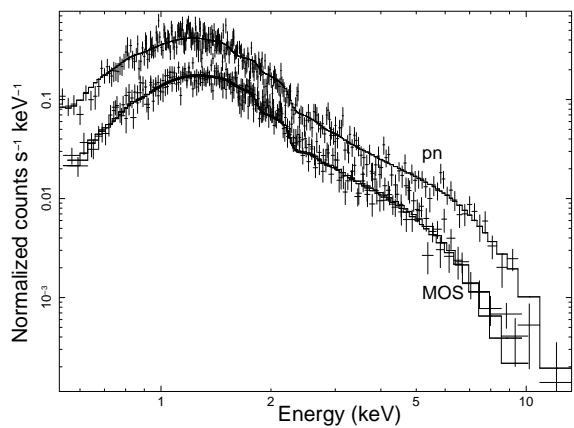}
\caption{The spectra of \textit{XMM-Newton} EPIC cameras, pn and the two MOS. The solid lines indicate the best fit to the data with a combined blackbody and power-law model.}
\label{f1}
 \end{center}
 
\end{figure}

To fit the spectra of the observations we used {\ttfamily XSPEC} (v~12.6.0). The spectra corresponding to the \textit{XMM-Newton} observation (of three EPIC cameras, the pn and the two MOS) are shown in Figure 1 and were fitted simultaneously with all parameters tied between the 3 detectors in order to provide the best constraints on the spectral parameters. The long effective exposure time ($\sim$ 17~ks) of this observation allows us to obtain the most accurate hydrogen column density, and it has good enough statistics to distinguish between fits using different models.\\
Firstly, we tried a power-law continuum model affected by absorption. The returned photon index was 2.74 $\pm$ 0.05 and the \Nh obtained was (0.53 $\pm$ 0.02)$\times 10^{22}$ cm$^{-2}$. However, this model led a poor fit (\chired =1.2 for 544 d.o.f.). Adding a blackbody as a soft component the fit improves notably (\chired =1.06 for 541 d.o.f.; see Fig.1). The parameters obtained with this model are \Nh (0.33 $\pm$ 0.03)$\times 10^{22}$ cm$^{-2}$, which is consistent with the value found by \citet{Kalberla2005} at the source position, a photon index of 1.7 $\pm$ 0.1, and a temperature (\textit{kT}) of 0.32 $\pm$ 0.02 keV. The soft component contributes nearly 30$\%$ of the 0.5-10 keV source flux. An f-test indicates a probability of 2.6 $\times 10^{-16}$ of achieving this level of improvement by chance.\\ 
The result is almost identical if we use a multicolor disc blackbody as the soft component. The \Nh was 0.37 $\pm$ 0.03 $\times 10^{22}$cm$^{-2}$; photon index was 1.6 $\pm$ 0.1 and a temperature at inner disc radius (T$_{in}$) was 0.45 $\pm$ 0.03 keV. 

The soft component cannot be constrained with \textit{Swift} data since their statistics are poorer, and neither with the \textit{RXTE} spectrum because it is not sensitive to energies below 2 keV. In a first attempt to study the evolution of the outburst, we calculate the X-ray colour using the \textit{Swift}/XRT data only to avoid calibration uncertainties between the different instruments. The color is defined as the ratio of counts between a hard band (2-10 keV) and a soft band (0.5-2 keV) and its values are shown in Fig.2~(c) and Fig.3~(b). We see that the spectrum becomes harder during the outburst, and it turns soft again when the outburst decays. This plot of the hardness ratio (HR) shows the spectral behaviour independently of the assumed spectral model.\\
To exclude the possibility that the observed spectral softening is due to pile-up (see also Section 2.3), we repeat the HR calculations using annular regions to exclude the photons coming from the center. We use annuli with an outer radius equal to the size of the circular region that was used previously (17 pixels; see Section 2.3), and three different sizes for the inner radius (7, 4 and 2 pixels). Our results using these different annuli are consistent with what is shown in Fig.2~(c) and Fig.3~(b), indicating that the softening is not related to pile-up.\\ 

In order to investigate the nature of this softening, we have carried out different spectral fits. First, we tested if the thermal component of the two component model varies. Since the poor statistics of the \textit{Swift} data do not permit fitting with a two component model, we made some assumptions. We fixed the \Nh and the photon index parameters with the values obtained in the \textit{XMM-Newton} fit. We took a power-law to represent the accretion flow, and the blackbody to represent the boundary layer. We fixed the power-law/blackbody ratio assuming that the relative efficiencies for the disc and the boundary layer will not vary, and we let the temperature vary freely. This was only possible for the two observations with highest count rates, observation 5 and 6 (see Table \ref{t1}). The temperatures resulting are 0.46 $\pm$ 0.06 keV and $0.56 ^{+ 0.05}_{-0.09 }$ keV, respectively. While the variation in temperature is not statistically significant, it is interesting to note that the data are consistent with the idea that only the blackbody temperature is varying.\\

To test the evolution along the outburst, we use a single power-law with absorption, since this is the only model that can fit all observations. The two component model is more unstable so the error estimates are much larger. The \Nh  was fixed to the value obtained from the {\it XMM-Newton} data (\Nh=0.53~$\times 10^{22}$~cm$^{-2}$), while the photon index and normalization components were left as free parameters. For the {\it Chandra} and the 6$^{th}$ {\it Swift} observation (Obs. 9 in Table 1) we used WebPIMMS to convert the obtained count rate into flux using the spectral parameters obtained in the {\it XMM-Newton} fitting.

For all cases, we calculated the absorbed and unabsorbed fluxes for both the 0.5-10 keV and the 2-10 keV energy ranges as well as the corresponding X-ray luminosities assuming a distance of 8 kpc, given the proximity of the source to the Galactic center. These results are reported in Table \ref{t2}. \\

\begin{figure}
\centering
\includegraphics{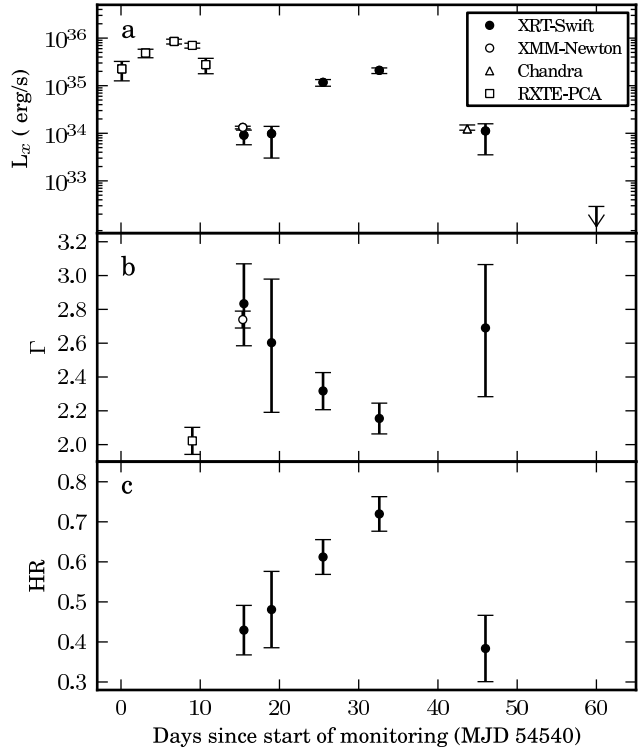}
\caption{(a) The light curve of XTE J1719-291 where the energy band is 2-10 keV. The first four white squares indicates \textit{RXTE} values that are taken from \citet{Markwardt2008}. (b) Photon index evolution. (c) Hardness ratio evolution (ratio of counts in the hard, 2-10 keV, and soft, 0.5-2 keV energy bands) using only the \textit{Swift} data.}
\label{f2}
\end{figure} 

The light curve (2-10 keV) is displayed in Fig. \ref{f2}~(a). In the plot we have included the four previous points from \textit{RXTE}/PCA bulge scans reported by \citet{Markwardt2008}. There are two peaks in the curve and the luminosity varies by $\sim$2 orders of magnitude. The peak luminosity value is 7 $\times$ $10^{35}$erg s$^{-1}$ on 2008 May 24. This low luminosity justifies a classification as a VFXT. The upper limit on the quiescent 2-10 keV luminosity inferred from the non-detection by \textit{Swift/XRT} on 2008 May 14 (Obs.9) is 2 $\times$ $10^{32}$erg s$^{-1}$. The outburst thus lasted at least ~46 days.\\ 

\begin{figure}
\centering
\includegraphics{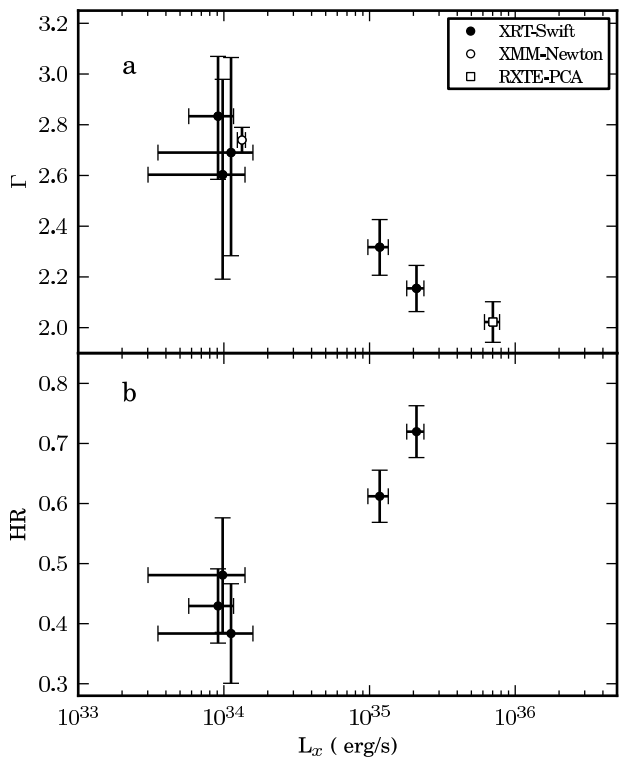}
\caption{Photon index (a) and hardness ratio using only \textit{Swift} data (b) (ratio of counts in the hard, 2-10 keV, and soft, 0.5-2 keV, energy bands) versus luminosity in the 2-10 keV energy band.}
\label{f3}
\end{figure}

In Fig. \ref{f2}~(b) the evolution of $\Gamma$ in time is plotted. We see variation of $\Gamma$ along the outburst, with values between 2 and 2.8. Comparing this figure with Fig. \ref {f2}~(a) it can be seen that $\Gamma$ increases with decreasing luminosity. In order to see this softening more clearly, we show a plot of $\Gamma$ versus $\lx$ in Fig. \ref{f3}~(a).

\subsection{Time-averaged accretion rate}

From the mean unabsorbed outburst flux we can estimate the average mass-accretion rate during outburst following the relation \Maccob=~$RL_\mathrm{acc}/GM$, where \textit{G} is the gravitational constant. $L_\mathrm{acc}$ is the 0.1-100~keV accretion luminosity which we estimate from the mean 2-10 keV unabsorbed outburst luminosity applying a bolometric correction factor of 3 \citep{Zand2007}. \textit{R} and \textit{M} are the radius and mass of the compact object, respectively. We obtain \Maccob~=~5.57~$\times$~$10^{-11}$~$\sol$~yr$^{-1}$ for a canonical neutron star (i.e. \textit{M}~=~1.4$\sol$, \textit{R}~=~10 km), and \Maccob~=~2.68~$\times$~$10^{-11}$~$\sol$~yr$^{-1}$ for a black hole (assuming \textit{M}~=~10$\sol$, \textit{R}~=~34 km). Once \Maccob is obtained, we determine the long-term averaged value using the relation \Macclong~=~ \Maccob~$\times$~$t_{\mathrm{ob}}$/$t_{\mathrm{rec}}$, where $t_{\mathrm{ob}}$ is the outburst duration and $t_{\mathrm{rec}}$ is the system's recurrence time, i.e., the sum of the outburst and quiescence time-scales. The factor $t_{\mathrm{ob}}$/$t_{\mathrm{rec}}$ is the duty cycle of the system.

For XTE J1719--291, $t_{\mathrm{ob}}$ is at least 46 days (see Fig.2). However, we do not know the quiescence time-scale because no other outbursts have been observed so far. We will assume a quiescence time ($t_{\mathrm{q}}$) of  at least 9 years, which is the time since RXTE-PCA has monitored this region during Galactic bulge scans (1999 February) till the discovery of XTE~J1719--291 on 2008 March. Taken this $t_{\mathrm{ob}}$, the duty cycle is $<$ 1.3$ \% $. This results in an estimated \Macclong~$ \lesssim $~7.7~$\times$~$10^{-13}$~$\sol$~yr$^{-1}$ for a neutron star compact object and \Macclong~$ \lesssim $~3.7~$\times$~$10^{-13}$~$\sol$~yr$^{-1}$ for a black hole. We note however, that outbursts could have been missed during the periods that the source could not be observed due to solar constraints.\\
We also have to consider is the fact that black holes systems might be radiatively inefficient at low accretion flows. Part of the generated accretion energy could be advected into the black hole or converted into jet power (e.g. \citealt{Blandford1999}; \citealt{Fender2003}; \citealt{Narayan2008}), therefore the estimation of \Macclong~ from the X-ray luminosity could be underestimated.\\

\section{Discussion}

We have presented \textit{RXTE}, \textit{Chandra}, \textit{XMM-Newton} and \textit{Swift} data analysis of the 2008 outburst of the newly discovered X-ray transient XTE J1719-291. The source was discovered on 2008 March 21 during \textit{RXTE}-PCA bulge scans and the outburst duration was at least 46 days (see Fig.2). The outburst light curve shows two peaks; the unabsorbed flux varies between (1.3-92) $\times$ $10^{-12}$~erg~cm$^{-2}$s$^{-1}$ (2-10 keV). Adopting a distance of 8~kpc, the inferred outburst peak luminosity is $\sim$ 7~$\times$ $10^{35}$~erg~s$^{-1}$ . This luminosity lies within the very faint X-ray regime, where $L^{peak}_{X}$(2-10 keV)~$<$ $10^{36}$~erg~s$^{-1}$. 
The nature of XTE~J1719-291 is unknown. An accreting white dwarf is very unlikely because these systems generally exhibit outburst peak luminosities below $10^{34}$~ergs~s$^{-1}$. Some classical novae have reached values of a few times $10^{34-35}$~erg~s$^{-1}$ for weeks to months, but none of them with a value as high as we find for XTE~J1719--291 \citep{Mukai2008}. Therefore the most likely origin of this X-ray luminosity value is a neutron star or black hole accreting system.\\

\subsection{X-ray spectral behaviour}

The high signal-to-noise of the \textit{XMM-Newton} spectra permits us to try different models to fit them. We found that a two component model, blackbody as a soft component and a power-law for the hard one,  could fit the spectra more accurately than a single component model. The best fit returned a temperature (kT) of 0.33 keV, a \Nh of 0.33~$\times~10^{22}$~cm$^{-2}$ and a photon index of 1.74. The blackbody component contributes ~30$\%$ of the total flux (0.5-10 keV). This soft component could be thermal emission from the surface of a neutron star or the boundary layer. One possible cause is accretion onto the neutron star at very low rates \citep{Zampieri1995}, but also can be incandescent thermal emission from the neutron star surface resulting from  deep crustal heating \citep{Brown1998}, which could be visible when the accretion disc becomes smaller. It was also possible to fit the spectra with a multicolor disc blackbody as the soft component (see Section 3). Therefore  the possibility that the emission comes from the accretion disc cannot be discarded. In fact, if the compact object is a black hole, the soft emission has to come from the disc.\\

We could detect the soft component robustly only in the \textit{XMM-Newton} data; the \textit{Swift} data lack sufficient signal-to-noise while \textit{RXTE}'s lower energy threshold of 2 keV is too high to allow detection of such a soft component. Therefore, in order to study the outburst spectral evolution, we fit all the data with the same model, this is a power-law continuum model affected by an equivalent hydrogen column. The photon index evolution shows a spectral softening, in other words, luminosity and photon index are anti-correlated. As we saw in section 3, we cannot rule out that the difference in the spectrum is produced by the blackbody soft component, i.e, the blackbody becomes stronger at lower $\lx$. In any case, the X-ray color diagram (Fig.3b) confirms the softening independently of the model used.\\
This behaviour differs from the bright transients systems, whose spectra evolve towards the low-hard state at the end of the outburst (see \citealt{Belloni2010}; \citealt{Klis2006}). However such softening towards even lower luminosities has been observed before in some black hole transients returning to quiescence from the hard-state. The photon index of XTE~J1650-500 softens from 1.66 to 1.93 in the hard state at X-ray luminosities down to $L_{X}$=1.5~$\times~10^{34}$~erg~s$^{-1}$ \citep{Tomsick2004}. XTE~J1550-564 and XTE~J1650-500 begin gradual softenings at low luminosities $L^{peak}_{X}$~$\lesssim$~$10^{36}$~erg~s$^{-1}$ \citep{kalemci2002}. Also \citet{Corbel2008} found that the photon index of V404 Cyg is softer in quiescence than in the hard state. This behaviour is consistent with the advection-dominated accretion flow (ADAF) model \citep{Esin1997} which predicts a gradual softening of the power-law photon index as the luminosity drops (see e.g. discussion in \citealt{Tomsick2004}). However, this is not always seen for all black holes in the last part of their outbursts. \citet{Jonker2009} did not find any evidence for this softening in the decay during the 2008 outburst of H~1743-322. It is worth pointing out that the black holes systems are fully described by a simple power-law model at these low luminosities, whereas we also detect a disc component in our \textit{XMM-Newton} spectrum. Therefore, the softening in our source also might be due to variations in the properties of the soft component.\\

On the other hand, we studied a thermal component evolution (see Section 3). We found hints that the temperature increases when the spectrum is brighter and harder. This could indicate that the softening is due to the variability of the temperature of the neutron star surface. According to Medvedev \& Narayan (2001) solutions, a hot optically thin region should be present in low L/L$_{\mathrm{edd}}$ neutron star systems, with a cooler boundary layer in the neutron star surface where the rotational energy is released.

\subsection{Optical counterpart and orbital period}

\begin{figure}
\begin{center}
\includegraphics[angle=-90,width=\columnwidth]{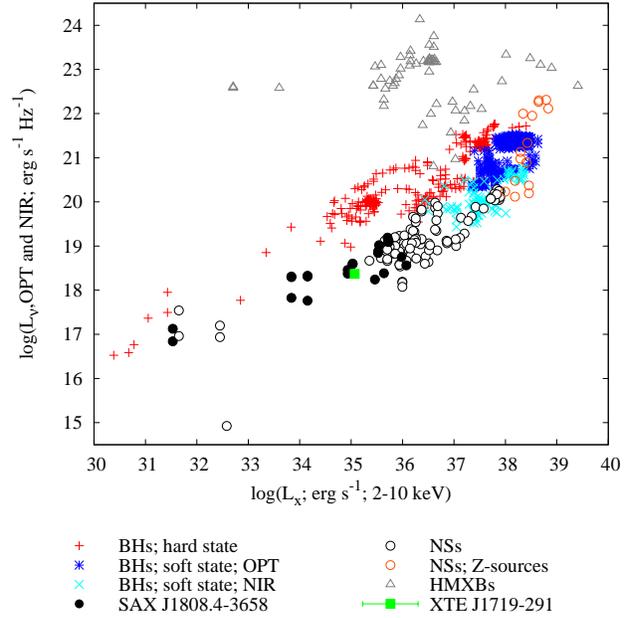}
\caption{The optical--X-ray luminosity diagram using the data of \citet{Russell2006,Russell2007a} for black holes (BHs), neutron stars (NSs) and HMXBs, with the data of XTE J1719--291 overplotted assuming a distance of 8 kpc to calculate the luminosities (green square). The NS transient SAX J1808.4--3658 (V4580 Sgr) is also indicated, as these data lie very close to XTE J1719--291 in the diagram.}
\label{optx}
 \end{center}
 
\end{figure}

An optical/NIR counterpart of XTE J1719--291 was first observed by \citet{Greiner2008} during an observation made on April 11, 2008. The counterpart was observed in several optical bands (i', r', g', z') with a magnitude between 22.3 and 23.0. The closest X-ray observation in time was made on April 9 by Swift (Obs. 5; Table \ref{t1}), with the source at a 0.5--10~keV luminosity of $~\sim ~3 \times 10^{35}$ erg s$^{-1}$ (see Table 2).  The optical counterpart was not detected in a subsequent observation made on May 4, 2008 when the X-ray luminosity was already below the sensitivity level of \textit{Swift}/XRT. Therefore the counterpart observed on April 11 is very likely optical emission from the accretion disc.

It was shown \citep{Russell2006,Russell2007a} that black holes and neutron stars occupy different regions of an optical--X-ray luminosity diagram when these transients are accreting at low luminosities ($L_{\rm X} \simlt 10^{36}$~erg~s$^{-1}$). At a given X-ray luminosity, a neutron star transient is typically $\sim 20$ times optically fainter than a black hole. We can therefore use the above quasi-simultaneous optical and X-ray luminosities of XTE J1719--291 to investigate the nature of its compact object by placing these data on this diagram. We estimate the de-reddened optical flux density adopting an extinction A$_{\rm i'}=2.11$, which has been calculated using the tabulated value reported by \citet{Schlegel1998} and by converting the value for the visual extinction of A$_{\rm V}=3.3$, as reported in \citet{Greiner2008}. To obtain the optical monochromatic luminosity $L_{\rm \nu, i'}$ \citep[flux density scaled to distance; see][]{Russell2006} and the X-ray 2--10 keV luminosity $L_{\rm X}$ we assume a distance of 8 kpc and an X-ray power law with photon index $\Gamma = 2.32$ (as measured for observation 5; Table \ref{t2}).

In Fig. \ref{optx} we plot the optical--X-ray luminosity diagram including data of all black holes, neutron stars and high-mass X-ray binaries (HMXBs) collected in \citet{Russell2006,Russell2007a,Russell2007b}, and overplot our data for XTE J1719--291. Errors are propagated from those quoted with the i'-band magnitude reported by \citet{Greiner2008} and the X-ray flux in Table \ref{t2}. At an assumed distance of 8 kpc, XTE J1719--291 lies amongst the other neutron star transients in the optical--X-ray luminosity diagram (Fig. \ref{optx}). At this X-ray luminosity, it is optically fainter than all the black holes in the sample, and $\sim 20$ times fainter in optical than a typical black hole. This provides evidence favouring a neutron star accretor in this VFXT, but this alone is no proof of the nature of the compact object; the source could indeed be an unusual black hole transient with a remarkably low optical/X-ray ratio.

The detection of an optical counterpart and the knowledge of the X-ray luminosity of the source are also useful to place some initial constraints on the orbital period of the system.  According to \citet{Paradijs1994}, the absolute visual magnitude of LMXBs correlates with the orbital period of the binary and the X-ray luminosity. If we assume M(i') $\approx$ M(V), the continuum spectral index is approximately flat ($F_{\nu} \propto \nu^{\sim 0.1}$), as may be expected for an LMXB disc at low luminosities slightly redder than a typical LMXB in outburst because the disc is probably cooler for this VFXT \citep{Hynes2005,Maitra2008} and again adopt A$_{\rm i'}=2.11$, we obtain M(i') and reach the following orbital period constraints.

If the compact object is a neutron star of $M=1.4\,M_{\odot}$ then Log$\left(\frac{P_{\rm orb}}{1\,\mathrm{hr}}\right)=-0.3^{+0.8}_{-0.7}$, whereas Log$\left(\frac{P_{\rm orb}}{1\,\mathrm{hr}}\right)=0.0^{+1.1}_{-0.4}$ in the case of a black hole of $M=10\,M_{\odot}$.  The orbital period is therefore in the range 0.4 $< P_{orb}< 12$ hr for a 10$M_{\odot}$ black hole, and 0.1 $< P_{\rm orb} < 3$ hr in case of a neutron star accretor (1$\sigma$ confidence intervals). If the system indeed hosts a neutron star, the binary is most likely to be compact or ultracompact since $P_{\rm orb} < 3$ hr, whereas this is not necessarily true for a black hole system, where $P_{orb}< 12$ hr.

\citet{Russell2006,Russell2007a} showed that the global empirical relations observed for a large sample of black holes and neutron stars can be approximated by the \citet{Paradijs1994} model; however the black holes are on average 10 times more luminous in optical than neutron stars. The scatter in optical monochromatic luminosity, defined as the mean of the differences between the data and the model, is $\pm 0.29$ dex for black holes and $\pm 0.36$ dex for neutron stars (both a factor of $\sim 2$). This scatter may be due to uncertainties in the distance, inclination, interstellar absorption and masses of each system, and possibly real, intrinsic effects. These relations and their scatter can be used to further constrain the likely value of the orbital period of XTE J1719--291 if it harbours either a neutron star or a black hole. If we again assume a neutron star of mass $M_1=1.4\,M_{\odot}$ and a companion of mass $M_2=0.6\,M_{\odot}$ \citep[typical values for the sample in][]{Russell2007a}, XTE J1719--291 would be consistent with the empirical relation for neutron stars if its orbital period is $P_{\rm orb} = 5.0^{+12.1}_{-3.5}$ hours. Alternatively, if the compact object is a black hole, XTE J1719--291 would only be consistent with the relation for black holes if its orbital period is $P_{\rm orb} = 0.08^{+0.13}_{-0.05}$ hours. This assumes a combined mass of the black hole and companion star of $M_1 + M_2 =10\,M_{\odot}$ \citep[typical for the sample in][]{Russell2006}. The significant differences between the orbital periods derived using the \citet{Paradijs1994} and \citet{Russell2006} relations result from the empirical systematic offset between black hole and neutron star sources found by the latter authors. The original \citet{Paradijs1994} relation was normalized to a collection of data containing two data points from black holes, and this systematic offset between black hole and neutron star accretors was only identified in a larger collection of sources using many data points from each source \citep[data from 15 black hole candidates and 19 neutron stars were used in][]{Russell2006,Russell2007a}.

These results favour a neutron star accretor in XTE J1719--291, with a likely orbital period of $1.5 \simlt P_{\rm orb} \simlt 17$ hr. It is also worth noting that XTE J1719--291 lies close to data of SAX J1808.4--3658 in the optical--X-ray luminosity diagram (Fig. \ref{optx}) which has an orbital period of 2.0 hours.

\subsection{Long-term average accretion rate}
We have calculated the long-term time-averaged accretion rate for both a neutron star and a black hole accretor (see Section 3.1). We find values of $10^{-13}$ to $10^{-12}$ $\sol$ yr$^{-1}$. These low accretion rates are difficult to explain with the current LMXB evolution models and it might be necessary to invoke exotic scenarios, such as neutron stars accreting from brown dwarfs or planetary companions (see \citealt{King2006}), although detail binary evolution calculations still need to be performed to support these conclusions. Other possibilities for these subluminous transients are the dissipation via radiatively inefficient flows of the accretion power for black holes (e.g. \citealt{Fender2003}; \citealt{Narayan2008}) or the \textquotedblleft propeller mechanism\textquotedblright~  for neutron stars, where only a fraction of  the mass transferred from the donor is accreted onto the neutron star (e.g.  \citealt{Illarionov1975};  \citealt{Alpar2001}; \citealt{Romanova2005} ).\\

\section*{Acknowledgments}

This work was supported by an ERC starting grant awarded to RW. AP and D.M.R acknowledge support from the Netherlands Organization for Scientific Research (NWO) Veni Fellowship

\bibliographystyle{mn2e}
\bibliography{bibliography}

\end{document}

%% file: definitions.tex
\def\lesssim{\mathrel{\hbox{\rlap{\hbox{\lower4pt\hbox{$\sim$}}}\hbox{$<$}}}}

\def\gtrsim{\mathrel{\hbox{\rlap{\hbox{\lower4pt\hbox{$\sim$}}}\hbox{$>$}}}}

\def\sol{~\mathrm{M}_\odot}
\def\lx{L_\mathrm{X}}

\def\Fx{$F_\mathrm{X}$}

\def\Nh{N$_{H}$~}

\def\Maccob{$\langle\dot{M}_{\mathrm{ob}}\rangle$}
\def\Macclong{$\langle\dot{M}_{\mathrm{long}}\rangle$}
\def\chired{$\chi^{2}_{\nu}$}

%% file: xtej1719.bbl
\begin{thebibliography}{}

\bibitem[\protect\citeauthoryear{{Alpar}}{{Alpar}}{2001}]{Alpar2001}
{Alpar} M.~A.,  2001, \apj, 554, 1245

\bibitem[\protect\citeauthoryear{{Belloni}}{{Belloni}}{2010}]{Belloni2010}
{Belloni} T.~M.,  2010, in {T.~Belloni} ed., Lecture Notes in Physics, Berlin
  Springer Verlag Vol.~794 of Lecture Notes in Physics, Berlin Springer Verlag,
  {States and Transitions in Black Hole Binaries}.
pp 53--+

\bibitem[\protect\citeauthoryear{{Blandford} \& {Begelman}}{{Blandford} \&
  {Begelman}}{1999}]{Blandford1999}
{Blandford} R.~D.,  {Begelman} M.~C.,  1999, \mnras, 303, L1

\bibitem[\protect\citeauthoryear{{Brown}, {Bildsten} \& {Rutledge}}{{Brown}
  et~al.}{1998}]{Brown1998}
{Brown} E.~F.,  {Bildsten} L.,    {Rutledge} R.~E.,  1998, \apjl, 504, L95+

\bibitem[\protect\citeauthoryear{{Chelovekov} \& {Grebenev}}{{Chelovekov} \&
  {Grebenev}}{2007}]{Chelovekov2007}
{Chelovekov} I.~V.,  {Grebenev} S.~A.,  2007, \astl, 33, 807

\bibitem[\protect\citeauthoryear{{Corbel}, {Koerding} \& {Kaaret}}{{Corbel}
  et~al.}{2008}]{Corbel2008}
{Corbel} S.,  {Koerding} E.,    {Kaaret} P.,  2008, \mnras, 389, 1697

\bibitem[\protect\citeauthoryear{{Cornelisse}, {Verbunt}, {in't Zand},
  {Kuulkers}, {Heise}, {Remillard}, {Cocchi}, {Natalucci}, {Bazzano} \&
  {Ubertini}}{{Cornelisse} et~al.}{2002}]{Cornelisse2002}
{Cornelisse} R.,  {Verbunt} F.,  {in't Zand} J.~J.~M.,  {Kuulkers} E.,  {Heise}
  J.,  {Remillard} R.~A.,  {Cocchi} M.,  {Natalucci} L.,  {Bazzano} A.,
  {Ubertini} P.,  2002, \aap, 392, 885

\bibitem[\protect\citeauthoryear{{Degenaar}, {Altamirano}, {Klein-Wolt} \&
  {Wijnands}}{{Degenaar} et~al.}{2008}]{Degenaar2008}
{Degenaar} N.,  {Altamirano} D.,  {Klein-Wolt} M.,    {Wijnands} R.,  2008, The
  Astronomer's Telegram, 1451, 1

\bibitem[\protect\citeauthoryear{{Degenaar}, {Altamirano} \&
  {Wijnands}}{{Degenaar} et~al.}{2008}]{Degenaar2008b}
{Degenaar} N.,  {Altamirano} D.,    {Wijnands} R.,  2008, The Astronomer's
  Telegram, 1467, 1

\bibitem[\protect\citeauthoryear{{Degenaar} \& {Wijnands}}{{Degenaar} \&
  {Wijnands}}{2008}]{Degenaar2008a}
{Degenaar} N.,  {Wijnands} R.,  2008, The Astronomer's Telegram, 1541, 1

\bibitem[\protect\citeauthoryear{{Degenaar} \& {Wijnands}}{{Degenaar} \&
  {Wijnands}}{2009}]{Degenaar2009}
{Degenaar} N.,  {Wijnands} R.,  2009, \aap, 495, 547

\bibitem[\protect\citeauthoryear{{Del Santo}, {Sidoli}, {Mereghetti},
  {Bazzano}, {Tarana} \& {Ubertini}}{{Del Santo} et~al.}{2007}]{DelSanto2007}
{Del Santo} M.,  {Sidoli} L.,  {Mereghetti} S.,  {Bazzano} A.,  {Tarana} A.,
  {Ubertini} P.,  2007, \aap, 468, L17

\bibitem[\protect\citeauthoryear{{Esin}, {McClintock} \& {Narayan}}{{Esin}
  et~al.}{1997}]{Esin1997}
{Esin} A.~A.,  {McClintock} J.~E.,    {Narayan} R.,  1997, \apj, 489, 865

\bibitem[\protect\citeauthoryear{{Fender}, {Gallo} \& {Jonker}}{{Fender}
  et~al.}{2003}]{Fender2003}
{Fender} R.~P.,  {Gallo} E.,    {Jonker} P.~G.,  2003, \mnras, 343, L99

\bibitem[\protect\citeauthoryear{{Gehrels}}{{Gehrels}}{1986}]{Gehrels1986}
{Gehrels} N.,  1986, \apj, 303, 336

\bibitem[\protect\citeauthoryear{{Greiner}, {Sala} \& {Kruehler}}{{Greiner}
  et~al.}{2008}]{Greiner2008}
{Greiner} J.,  {Sala} G.,    {Kruehler} T.,  2008, The Astronomer's Telegram,
  1577, 1

\bibitem[\protect\citeauthoryear{{Hynes}}{{Hynes}}{2005}]{Hynes2005}
{Hynes} R.~I.,  2005, \apj, 623, 1026

\bibitem[\protect\citeauthoryear{{Ibrahim}, {Markwardt}, {Swank}, {Ransom},
  {Roberts}, {Kaspi}, {Woods}, {Safi-Harb}, {Balman}, {Parke}, {Kouveliotou},
  {Hurley} \& {Cline}}{{Ibrahim} et~al.}{2004}]{Ibrahim2004}
{Ibrahim} A.~I.,  {Markwardt} C.~B.,  {Swank} J.~H.,  {Ransom} S.,  {Roberts}
  M.,  {Kaspi} V.,  {Woods} P.~M.,  {Safi-Harb} S.,  {Balman} S.,  {Parke}
  W.~C.,  {Kouveliotou} C.,  {Hurley} K.,    {Cline} T.,  2004, \apjl, 609, L21

\bibitem[\protect\citeauthoryear{{Illarionov} \& {Sunyaev}}{{Illarionov} \&
  {Sunyaev}}{1975}]{Illarionov1975}
{Illarionov} A.~F.,  {Sunyaev} R.~A.,  1975, \aap, 39, 185

\bibitem[\protect\citeauthoryear{{in't Zand}, {Jonker} \& {Markwardt}}{{in't
  Zand} et~al.}{2007}]{Zand2007}
{in't Zand} J.~J.~M.,  {Jonker} P.~G.,    {Markwardt} C.~B.,  2007, \aap, 465,
  953

\bibitem[\protect\citeauthoryear{{Jonker}, {Miller-Jones}, {Homan}, {Gallo},
  {Rupen}, {Tomsick}, {Fender}, {Kaaret}, {Steeghs}, {Torres}, {Wijnands},
  {Markoff} \& {Lewin}}{{Jonker} et~al.}{2009}]{Jonker2009}
{Jonker} P.~G.,  {Miller-Jones} J.,  {Homan} J.,  {Gallo} E.,  {Rupen} M.,
  {Tomsick} J.,  {Fender} R.~P.,  {Kaaret} P.,  {Steeghs} D.~T.~H.,  {Torres}
  M.~A.~P.,  {Wijnands} R.,  {Markoff} S.,    {Lewin} W.~H.~G.,  2009, \mnras,
  pp 1618--+

\bibitem[\protect\citeauthoryear{{Kalberla}, {Burton}, {Hartmann}, {Arnal},
  {Bajaja}, {Morras} \& {P{\"o}ppel}}{{Kalberla} et~al.}{2005}]{Kalberla2005}
{Kalberla} P.~M.~W.,  {Burton} W.~B.,  {Hartmann} D.,  {Arnal} E.~M.,  {Bajaja}
  E.,  {Morras} R.,    {P{\"o}ppel} W.~G.~L.,  2005, \aap, 440, 775

\bibitem[\protect\citeauthoryear{{Kalemci}}{{Kalemci}}{2002}]{kalemci2002}
{Kalemci} E.,  2002, PhD thesis, UNIVERSITY OF CALIFORNIA, SAN DIEGO

\bibitem[\protect\citeauthoryear{{King} \& {Wijnands}}{{King} \&
  {Wijnands}}{2006}]{King2006}
{King} A.~R.,  {Wijnands} R.,  2006, \mnras, 366, L31

\bibitem[\protect\citeauthoryear{{Lasota}}{{Lasota}}{2001}]{Lasota2001}
{Lasota} J.,  2001, New Astronomy Review, 45, 449

\bibitem[\protect\citeauthoryear{{Maitra} \& {Bailyn}}{{Maitra} \&
  {Bailyn}}{2008}]{Maitra2008}
{Maitra} D.,  {Bailyn} C.~D.,  2008, \apj, 688, 537

\bibitem[\protect\citeauthoryear{{Markwardt} \& {Swank}}{{Markwardt} \&
  {Swank}}{2008}]{Markwardt2008}
{Markwardt} C.~B.,  {Swank} J.~H.,  2008, The Astronomer's Telegram, 1442, 1

\bibitem[\protect\citeauthoryear{{Masetti}, {Landi}, {Pretorius}, {Sguera},
  {Bird}, {Perri}, {Charles}, {Kennea}, {Malizia} \& {Ubertini}}{{Masetti}
  et~al.}{2007}]{Masetti2007}
{Masetti} N.,  {Landi} R.,  {Pretorius} M.~L.,  {Sguera} V.,  {Bird} A.~J.,
  {Perri} M.,  {Charles} P.~A.,  {Kennea} J.~A.,  {Malizia} A.,    {Ubertini}
  P.,  2007, \aap, 470, 331

\bibitem[\protect\citeauthoryear{{Mukai}, {Orio} \& {Della Valle}}{{Mukai}
  et~al.}{2008}]{Mukai2008}
{Mukai} K.,  {Orio} M.,    {Della Valle} M.,  2008, \apj, 677, 1248

\bibitem[\protect\citeauthoryear{{Muno}, {Gaensler}, {Clark}, {de Grijs},
  {Pooley}, {Stevens} \& {Portegies Zwart}}{{Muno} et~al.}{2007}]{Muno2007}
{Muno} M.~P.,  {Gaensler} B.~M.,  {Clark} J.~S.,  {de Grijs} R.,  {Pooley} D.,
  {Stevens} I.~R.,    {Portegies Zwart} S.~F.,  2007, \mnras, 378, L44

\bibitem[\protect\citeauthoryear{{Narayan} \& {McClintock}}{{Narayan} \&
  {McClintock}}{2008}]{Narayan2008}
{Narayan} R.,  {McClintock} J.~E.,  2008, \nar, 51, 733

\bibitem[\protect\citeauthoryear{{Okazaki} \& {Negueruela}}{{Okazaki} \&
  {Negueruela}}{2001}]{Okazaki2001}
{Okazaki} A.~T.,  {Negueruela} I.,  2001, \aap, 377, 161

\bibitem[\protect\citeauthoryear{{Romanova}, {Ustyugova}, {Koldoba} \&
  {Lovelace}}{{Romanova} et~al.}{2005}]{Romanova2005}
{Romanova} M.~M.,  {Ustyugova} G.~V.,  {Koldoba} A.~V.,    {Lovelace} R.~V.~E.,
   2005, \apjl, 635, L165

\bibitem[\protect\citeauthoryear{{Russell}, {Fender}, {Hynes}, {Brocksopp},
  {Homan}, {Jonker} \& {Buxton}}{{Russell} et~al.}{2006}]{Russell2006}
{Russell} D.~M.,  {Fender} R.~P.,  {Hynes} R.~I.,  {Brocksopp} C.,  {Homan} J.,
   {Jonker} P.~G.,    {Buxton} M.~M.,  2006, \mnras, 371, 1334

\bibitem[\protect\citeauthoryear{{Russell}, {Fender} \& {Jonker}}{{Russell}
  et~al.}{2007a}]{Russell2007a}
{Russell} D.~M.,  {Fender} R.~P.,    {Jonker} P.~G.,  2007a, \mnras, 379, 1108

\bibitem[\protect\citeauthoryear{{Russell}, {Maccarone}, {K{\"o}rding} \&
  {Homan}}{{Russell} et~al.}{2007b}]{Russell2007b}
{Russell} D.~M.,  {Maccarone} T.~J.,  {K{\"o}rding} E.~G.,    {Homan} J.,
  2007b, \mnras, 379, 1401

\bibitem[\protect\citeauthoryear{{Schlegel}, {Finkbeiner} \&
  {Davis}}{{Schlegel} et~al.}{1998}]{Schlegel1998}
{Schlegel} D.~J.,  {Finkbeiner} D.~P.,    {Davis} M.,  1998, \apj, 500, 525

\bibitem[\protect\citeauthoryear{{Tomsick}, {Kalemci} \& {Kaaret}}{{Tomsick}
  et~al.}{2004}]{Tomsick2004}
{Tomsick} J.~A.,  {Kalemci} E.,    {Kaaret} P.,  2004, \apj, 601, 439

\bibitem[\protect\citeauthoryear{{van der Klis}}{{van der
  Klis}}{2006}]{Klis2006}
{van der Klis} M.,  2006, \adspr, 38, 2675

\bibitem[\protect\citeauthoryear{{van Paradijs} \& {McClintock}}{{van Paradijs}
  \& {McClintock}}{1994}]{Paradijs1994}
{van Paradijs} J.,  {McClintock} J.~E.,  1994, \aap, 290, 133

\bibitem[\protect\citeauthoryear{{Werner}, {in't Zand}, {Natalucci},
  {Markwardt}, {Cornelisse}, {Bazzano}, {Cocchi}, {Heise} \&
  {Ubertini}}{{Werner} et~al.}{2004}]{Werner2004}
{Werner} N.,  {in't Zand} J.~J.~M.,  {Natalucci} L.,  {Markwardt} C.~B.,
  {Cornelisse} R.,  {Bazzano} A.,  {Cocchi} M.,  {Heise} J.,    {Ubertini} P.,
  2004, \aap, 416, 311

\bibitem[\protect\citeauthoryear{{Wijnands}}{{Wijnands}}{2008}]{Wijnands2008}
{Wijnands} R.,  2008, in {Bandyopadhyay} R.~M.,  {Wachter} S.,  {Gelino} D.,
  {Gelino} C.~R.,  eds, A Population Explosion: The Nature \& Evolution of
  X-ray Binaries in Diverse Environments Vol.~1010 of American Institute of
  Physics Conference Series, {Enigmatic Sub-luminous Accreting Neutron Stars in
  our Galaxy}.
pp 382--386

\bibitem[\protect\citeauthoryear{{Wijnands}, {in't Zand}, {Rupen}, {Maccarone},
  {Homan}, {Cornelisse}, {Fender}, {Grindlay}, {van der Klis}, {Kuulkers},
  {Markwardt}, {Miller-Jones} \& {Wang}}{{Wijnands}
  et~al.}{2006}]{wijnands2006}
{Wijnands} R.,  {in't Zand} J.~J.~M.,  {Rupen} M.,  {Maccarone} T.,  {Homan}
  J.,  {Cornelisse} R.,  {Fender} R.,  {Grindlay} J.,  {van der Klis} M.,
  {Kuulkers} E.,  {Markwardt} C.~B.,  {Miller-Jones} J.~C.~A.,    {Wang} Q.~D.,
   2006, \aap, 449, 1117

\bibitem[\protect\citeauthoryear{{Wijnands}, {Miller} \& {Wang}}{{Wijnands}
  et~al.}{2002}]{Wijnands2002}
{Wijnands} R.,  {Miller} J.~M.,    {Wang} Q.~D.,  2002, \apj, 579, 422

\bibitem[\protect\citeauthoryear{{Zampieri}, {Turolla}, {Zane} \&
  {Treves}}{{Zampieri} et~al.}{1995}]{Zampieri1995}
{Zampieri} L.,  {Turolla} R.,  {Zane} S.,    {Treves} A.,  1995, \apj, 439, 849

\end{thebibliography}
